\documentclass[12pt]{iopart}
\usepackage{iopams}
\usepackage{epsfig}

\newlength{\mylen}
\newcommand{\bea}{\begin{eqnarray}}
\newcommand{\eea}{\end{eqnarray}}

\newcommand{\vect}[1]{\mathbf{#1}}

\newcommand{\ef}{\varepsilon_F}

\newcommand{\Ezp}{E_{z,0^+}}

\newcommand{\Ep}{E_{||,0}}
\newcommand{\ro}{r_{0,{\rm ref}}}
\newcommand{\agt}{\gtrsim}
\newcommand{\alt}{\lesssim}

\begin{document}

\title{Attractions between charged colloids at water interfaces}

\author{M. Oettel\dag \footnote[3]
  {To whom correspondence should be addressed (oettel@mf.mpg.de)}, 
        A. Dom\'\i nguez\ddag, and 
        S. Dietrich\dag}
%\author{A. Dom\'\i nguez\ddag}
%\author{S. Dietrich\dag}

\address{\dag\
Max--Planck--Institut f\"ur Metallforschung, Heisenbergstr. 3, D-70569 Stuttgart, \\
Institut f\"ur Theoretische und Angewandte Physik, Universit\"at Stuttgart,
 Pfaffenwaldring 57, D-70569 Stuttgart, Germany}
\address{\ddag\
F\'\i sica Te\'orica, Universidad de Sevilla, Apdo.~1065, E-41080
  Sevilla, Spain}

\begin{abstract}
 The effective  potential  between charged colloids trapped at water interfaces
 is analyzed. It consists of a repulsive electrostatic and an attractive 
 capillary part which asymptotically both show  dipole--like behavior. For 
 sufficiently large colloid 
 charges, the capillary attraction dominates at large separations.
 The total effective potential exhibits a minimum at intermediate
 separations if the Debye screening length of water and the colloid radius
 are of comparable size.   
\end{abstract}


\submitto{\JPCM}

%\maketitle

\section*{}

In view of various basic and applied issues such as the study of two--dimensional melting
\cite{Pie80}, investigations of mesoscale structure formation \cite{Joa01}
or engineering of colloidal crystals on spherical surfaces \cite{Din02}, 
the self--assembly of 
sub-$\mu$m colloidal particles at water--air or water--oil interfaces
has gained much interest in recent years. These particles are trapped at the interface
 if water wets the colloid only partially. 
This configuration is stable against thermal fluctuations. It appears to be even 
the global equilibrium state,
because it is observed experimentally that the colloids immersed in the bulk phases
are attracted towards the interface \cite{Pie80}. 
For charge--stabilized colloids at interfaces, the {\em repulsive} part of their  mutual 
interaction is well understood and resembles a dipole--dipole interaction at large
separations. This asymptotic interaction is caused by charges at the 
colloid--water 
interface \cite{Hur85} or by isolated charges at the colloid--air (or oil) 
interface \cite{Ave00a}.
Nonetheless, charged colloids at interfaces exhibit also {\em attractions} far 
beyond the range of van--der--Waals forces.  According to
Refs.~\cite{Ghe97,Ghe01,Gar98a,Gar98b,Sta00,Que01}, polystyrene spheres
(radii $R=0.25\dots 2.5$ $\mu$m)
 on flat water--air interfaces  using deionized water exhibit
spontaneous formation
of complicated metastable mesostructures. They are consistent with the presence
of an attractive, secondary
minimum in the effective intercolloidal potential at separations $d/R\approx 3\dots 20$ with
a depth of a few $k_B T$. In Ref.~\cite{Nik02}, PMMA spheres with radius $R=0.75$ 
$\mu$m were trapped
at the interface of water droplets immersed in oil.
Here, the secondary minimum has been detected at a
separation $d/R=7.6$ and is reported to be  surprisingly steep.
The tentative explanation of these
findings given in Ref.~\cite{Nik02} invokes an analogue of long--ranged
flotation or capilllary forces which decay $\propto 1/d$. 
This interpretation  was criticized in 
Refs.~\cite{Meg03,For04} 
which both concluded that possible capillary forces in
this system are much shorter ranged, i.e.,   $\propto d^{-7}$, 
but the authors of these references disagree with respect to
the sign of these shorter--ranged forces. Recently we have shown \cite{Oet05}
quite generally that long--ranged flotation--like forces $\propto 1/d$ can only arise
in mechanically non--isolated systems. For isolated systems the capillary 
force is indeed much shorter--ranged and within a superposition approximation the 
power law discussed in Ref.~\cite{For04} is found. 
For the experiment involving the interface of a mesoscopic 
droplet \cite{Nik02} mechanical isolation may indeed be weakly violated and thus small 
flotation--like forces can appear. Their interplay with capillary forces arising from
the droplet curvature is not clear yet and is currently under scrutiny \cite{Dom05}.
On the other hand, for the experiments performed on flat interfaces 
\cite{Ghe97,Ghe01,Gar98a,Gar98b,Sta00,Que01}, mechanical isolation 
holds if there is no external  electric field present and thus for them 
flotation--like forces 
cannot give rise to the observed attractions.
%
%There are two other 
%Among other attempts to explain the nature of this colloidal pattern 
%formation, we can mention two. (i) In 
%Ref.~\cite{FMMH04} the appearance of mesoscopic 
%colloid patterns at the water--air interface (see, e.g., Ref.~\cite{Ghe97})
%has been attributed to oil contaminations of the interface which may form patches 
%of the same mesoscopic size in which the colloids are trapped. 
%(ii) According to Ref.~\cite{Sta00}, pronounced colloidal surface roughness may lead 
%to anisotropic capillary forces with an asymptotically quadrupolar decay 
%$\propto d^{-5}$.
Out of other attempts to explain the nature of this colloidal pattern
formation we mention Ref.~\cite{FMMH04}, in which this is attributed 
to oil contaminations of the interface, and Ref.~\cite{Sta00}, in which
colloidal roughness is proposed as a source of attractive capillary interactions.
At present, these attempts to explain the observed
colloidal patterns at interfaces are only of qualitative nature.

Thus a theoretically sound mechanism for the appearance of an attractive
minimum in the intercolloidal potential at large separations has not been found yet.
Here we analyze the interaction between colloids at interfaces within the approach 
developed in Ref.~\cite{Oet05} for a mechanically isolated system 
and we provide conditions for an asymptotically attractive intercolloidal
potential and for the appearance of such a minimum.

In going beyond the %previously analyzed 
superposition approximation studied in Ref.~\cite{Oet05}, we derive the full capillary interaction potential
between two colloids as a functional of a general stress field acting on the interface. This 
capillary potential is studied for two cases: (i) the Debye screening
length of water, $\kappa^{-1}$, is much smaller than the colloid radius $R$, and (ii)
$R$ and $\kappa^{-1}$ are of comparable size. For a sufficiently high charge on the
colloids, in both cases
the ensuing capillary attraction turns out to be asymptotically stronger than the
direct repulsion.
Moreover, in case (ii) a {\em minimum} in the total (repulsive plus capillary)
intercolloidal potential is found at intermediate separation
$\kappa d_{\rm min} \agt 10$.        

We consider two spherical colloids $\alpha=1,2$ trapped at a deformed  interface (meniscus) with vertical
coordinate  $z=\hat u(\vect r=(x,y))$. We denote $\hat h_\alpha$ as the vertical 
coordinate of 
the center of colloid $\alpha$ and $\vect r_\alpha$ its lateral position so that 
$d=|\vect r_1-\vect r_2|$.
We define a reference configuration (with respect to
which free energy differences are measured) by a flat interface  
$u=0$ and $h_{\alpha,\rm ref}=-R\cos\theta$. Here, $\theta$ is Young's angle and thus
 in the reference configuration the colloids are vertically positioned such that Young's law
holds at the horizontal three--phase contact circle with radius $\ro=R\sin\theta$.
The corresponding free energy is given by \cite{Oet05}
\bea
  \fl
  \label{eq:F2}
  {\hat {\cal F}} &= &
  \gamma \int_{S_{\rm men}} \!\!\!\!\!\! d^2 r \;
  \left[  \frac{|\nabla \hat{u}|^2}{2} + \frac{\hat{u}^2}{2 \lambda^2} -
    \frac{\hat{\Pi}}{\gamma} \, \hat{u} \right] + 
% \\ & &
   \sum_{\alpha=1,2} \left\{
    \frac{\gamma}{2 r_{0, \rm ref}} \oint_{\partial S_\alpha} \!\!\! d\ell \; [\Delta \hat{h}_\alpha - \hat{u}]^2
    - \hat{F}_\alpha \Delta \hat{h}_\alpha \right\} . 
% \nonumber
\eea
The first line of Eq.~(\ref{eq:F2}) comprises free energy differences associated
with the change in meniscus area, in meniscus gravitational energy ($\gamma$ is the
water--air surface tension and $\lambda$
is the capillary length) and with forces on the meniscus; the stress 
$\hat{\Pi}$ denotes 
the vertical force per unit area on the meniscus in the reference configuration.
The first term in the second line takes into account the changes in water--colloid and 
air--colloid surface energies; $\Delta \hat{h}_\alpha$ is the difference in 
colloid center position with respect to the reference configuration. 
The second term describes the energy difference if colloid $\alpha$ is shifted
by the force $\hat{F}_\alpha$ (which is evaluated in the
reference configuration). $S_\alpha$ is the circular disk of radius $\ro$ delimited by
the three--phase contact line $\partial S_\alpha$ formed  on the colloid $\alpha$ 
by the fluid interface in the reference configuration (traced counterclockwise), 
and $S_{\rm men} =
\mathbb{R}^2\backslash(S_1 \bigcup S_2)$. This expression for the free energy is 
valid as long as the 
deviations from the reference configuration are small: $|\hat u|/\ro,|\nabla \hat{u}| \ll 1$.
A sufficient condition for this is
$|\varepsilon_{\hat F}| \ll 1$, with 
$\varepsilon_{\hat F}:=-\hat{F}_\alpha/\; 2\pi\gamma\ro$.
Mechanical isolation implies $2 {\hat F_\alpha} = 
- \int_{S_{\rm men}} \!\!\!\!\!\! d^2r\, \hat \Pi$, 
i.e., the forces on the colloids are balanced by the force on the meniscus \cite{Oet05}.
The meniscus--induced effective potential between the colloids is defined as
%\bea
% \label{eq:vmen}
$V_{\rm men}(d) = \hat {\cal F}(\hat u,\hat h_\alpha|_{\rm eq};d) - 
\hat {\cal F}(\hat u,\hat h_\alpha|_{\rm eq};d \to \infty)$. 
%\eea
The equilibrium free energy is found by minimizing Eq.~(\ref{eq:F2}) with respect to
$\hat u$ and $\hat h_\alpha$. The equilibrium meniscus shape for
the two colloids being infinitely apart is given by the superposition 
$\hat u(d\to \infty) = u_1 + u_2$ of the single colloid menisci, obtained in the
presence of the stress field $\hat \Pi(d\to \infty) = \Pi_1+ \Pi_2$.
Here, $u_\alpha = u(|\vect r-\vect r_\alpha|)$ is the equilibrium meniscus around
{\em one} colloid 
and likewise $\Pi_\alpha = \Pi(|\vect r-\vect r_\alpha|)$ is the stress field on the 
interface caused by a {\em single} colloid. Furthermore we define
the single-colloid quantity $\ef = \int_{\mathbb{R}^2\backslash S_1} 
 \!\!d^2r\, \Pi /\; 2\pi\gamma\ro$. For finite $d$, the stress field can be 
written as $\hat \Pi(d) = \Pi_1+ \Pi_2 + 2\,\Pi_{\rm m}$ 
and likewise we decompose $\hat u(d) = u_1+u_2+ u_{\rm m}$. 
Minimization of Eq.~(\ref{eq:F2}) yields
\bea
 \label{eq:um}
 \nabla^2 u_{\rm m} - \frac{u_{\rm m}}{\lambda^2} = - \frac{2\,\Pi_{\rm m}}{\gamma}\;,
\eea
with the boundary condition $u_{\rm m}=0$ for
$\vect r \to \infty$, and at $\vect r \in \partial S_\alpha$ ($\beta \not = \alpha$):
\bea
 \label{eq:umb}
 \frac{\partial(u_{\rm m}+u_\beta)}{\partial n_\alpha} = \varepsilon_{\hat F}-\ef +
  \frac{u_{\rm m}+u_\beta-\langle u_{\rm m}+u_\beta \rangle_\alpha}{\ro} \;,
\eea
with $2\pi\ro\,\langle \cdot \rangle_\alpha \equiv 
\oint_{\partial S_\alpha}(\cdot) d\ell$.
The superposition approximation entails $\Pi_{\rm m}=u_{\rm m}=0$ for all $d$ (i.e., $\varepsilon_{\hat F}=\ef$), and
was analyzed in Ref.~\cite{Oet05}. To identify the corrections to it, we
write $V_{\rm men}=V_{\rm sup}[\Pi] + V_{\rm m}[\hat \Pi]$ 
and obtain after some algebra
(in the following we consider $\lambda \to \infty$; this limit can be safely
taken in the case of mechanical isolation \cite{Oet05})
\bea
  V_{\rm sup} &=& \int_{S_1} \!\!\!d^2r\,\Pi_2u_2 -
           \int_{S_{\rm men}}\!\!\!\!\!\!\!\!\!\! d^2r\,\Pi_1u_2 +
            2\pi\gamma\ro\,\ef\,\langle u_2\rangle_1\;, \nonumber \\
  \label{eq:vm} 
% \\
  V_{\rm m} &=& -\int_{S_{\rm men}} \!\!\!\!\!\!\!\!\!\!d^2r\,
          (\Pi_2 u_{\rm m} +
         2\, \Pi_{\rm m} u_2 + \Pi_{\rm m} u_{\rm m}) + 2\pi\gamma\ro^2 
% \nonumber 
 \\
%   & \times & \left[(\ef^2-\varepsilon_{\hat F}^2)  + 
   &  & \times \left[(\ef^2-\varepsilon_{\hat F}^2)  + 
  \varepsilon_{\hat F}\frac{\langle u_{\rm m}\rangle_1}{\ro} - 
     (\ef-\varepsilon_{\hat F}) \frac{\langle u_1 +u_2\rangle_1}{\ro} \right]\;.\nonumber 
\eea
In the following, we apply this general expression for the capillary potential
within two electrostatic models which provide explicit expressions for $\Pi$ and
$\Pi_{\rm m}$.  Generically, the water
phase contains screening ions which lead to a finite Debye screening 
length $\kappa^{-1}=(\epsilon_2/\; 8\pi\beta c_0 e^2)^{1/2}$ 
where $c_0$ is the concentration of monovalent ions, $e$ is the elementary charge, and 
$\beta^{-1}=k_B T$. We denote by
$\epsilon_1$, $\epsilon_2$, $\epsilon_{\rm c}$ the permittivities
of air, water, and the colloid, respectively, using Gau{\ss} units.

{\bf 1. $\kappa \ro \gg 1$.} The electrostatics of the single--colloid configuration has been analyzed in 
Ref.~\cite{Kra04} with the result that only charges at the colloid--air
interface generate a stress on the interface. The potential 
$\Phi_0$ at the air--water interface is small so that
the interface resembles a perfect conductor. According to Ref.~\cite{Kra04},
$\Pi \approx (\epsilon_1/\,8\pi)
\Ezp^2$, where $\Ezp$ denotes the $z$--component of the electric field right above
the interface, leading to the following
robust parametrization of the corresponding numerical results:
\bea
 \Pi(r) = \frac{\gamma\ef}{\ro} b(\mu)\left( \frac{r}{\ro} -1 \right)^{\mu-1} 
      \left( \frac{r}{\ro}\right)^{-\mu-5} \;,
\eea 
where $\ef > 0$, $b=\mu(\mu+1)(\mu+2)(\mu+3)/6$ and $\mu \in(0,1)$ is a fitting parameter
depending on $\epsilon_{\rm c}$ and the contact angle $\theta$. $\Pi$ has an 
integrable divergence as $r\to \ro$ and rapidly (i.e., $r\agt 2\ro$) 
reaches its
asymptotic dipole behavior $\propto r^{-6}$. The dipole asymptotics 
renders the repulsive part of the intercolloidal potential $V_{\rm rep}$
for large $d$,
\bea
 \label{eq:vrep_mod1}
 V_{\rm rep} = 4\pi\gamma\ro^2\,\ef b(\mu) \left(\frac{\ro}{d}\right)^3 \;,
\eea
\nopagebreak
and $\Pi_{\rm m} = (\Pi_1\Pi_2)^{1/2}$ because %for large $d$ 
the electric field of
the two--colloid configuration is the superposition of the electric fields in the
single--colloid configurations. 
This stress is strongly peaked around the colloid centers and the main
contribution in Eqs.~(\ref{eq:vm}) stems from the regions around the
colloids as $d\to\infty$. Therefore, in order to obtain
$V_{\rm men}$ to leading order in $1/d$ one can employ the
approximation $\Pi_{\rm m} \approx
\Pi^{1/2}(d)(\Pi_1^{1/2}+\Pi_2^{1/2})$ to solve Eqs.~(\ref{eq:um}) and
(\ref{eq:umb}) and to evaluate the integrals in Eq.~(\ref{eq:vm}):
\bea
 \label{eq:vmen_mod1}
 V_{\rm men}(d) \approx -2\pi\gamma \ro^2\, \ef^2\, 
 \left(\frac{\ro}{d}\right)^3 \,
  \frac{8b(\mu)}{\mu+1}\,[1+ M(\mu)]\;.
\eea
Here, the function $M(\mu)$ is given by a lengthy analytical expression; 
it increases almost linearly for $\mu \in (0,1)$
with $M(0)=0$ and $M(1)=1/5$.
The leading asymptotic behavior $\propto d^{-3}$ stems from $V_{\rm m}$ (the 
superposition
approximation predicts $V_{\rm sup} \propto d^{-6}$ \cite{Oet05}). 
Repulsive and capillary force
decay with the same power but with opposite signs of the amplitudes. 
Hence, the
total intercolloidal potential will be attractive if 
$\ef > \varepsilon_{F,{\rm crit}}= (\mu+1)/ \,4[1+M(\mu)]$ with
$1/4<\varepsilon_{F,{\rm crit}}<5/12$ for $0<\mu<1$. 
The appearance of capillary attractions 
thus depends sensitively on the colloidal charge via $\ef$.
These critical values for $\ef$ 
are at the limit of applicability of this model: The maximum gradient and
the maximum deviation for the meniscus occur at the three--phase contact line and are given to leading order in $1/d$ by the single--colloid solution:
$|\nabla \hat{u}| \approx \ef$, $\hat{u}/\ro\approx -\mu\ef/4$.
Experiments which  are performed in the limit
$\kappa\ro \gg 1$ \cite{Ave00a,Kra04} estimate charge densities on the 
colloid--air surface compatible with $0< \ef \ll 1$ and thus the electrostatic
repulsion would always be stronger than the capillary attraction. Even if
$\ef$ were large enough such that the capillary attraction would dominate, 
Eqs.~(\ref{eq:vrep_mod1}) and (\ref{eq:vmen_mod1}) would not render 
a minimum in the total potential at {\em intermediate}
separations. However, the possibility for the occurence of such a minimum arises 
outside the regime $\kappa\ro \gg 1$.

{\bf 2. $\kappa\ro \alt 1$.} In this regime, $\kappa^{-1}$ 
provides an additional  length scale 
which leads to interesting crossover phenomena 
and the charges on the colloid--water surface are not completely screened in the 
range of separations of interest.
The stress is given by
\bea
\label{eq:stress}
 \Pi = \frac{\epsilon_1}{8\pi}\left( 1- \frac{\epsilon_1}{\epsilon_2}\right) \left[ \Ezp^2   +
   \frac{\epsilon_2}{\epsilon_1}\, \Ep^2 \right] + p_{\rm osm}\;.
\eea
Here, $\Ezp$ and $\Ep$ are the perpendicular and the parallel component of the 
electric field at the interface on the air side. 
The osmotic pressure $p_{\rm osm}=\beta^{-1} \Delta n$ is generated by the excess ion
number density $\Delta n$ at the interface on the water side.
In order to solve the electrostatic problem, we introduce two
simplifications: the Debye-H\"uckel approximation, and the
point--charge approximation, by which the total charge $q$ of the
colloid is concentrated at the center of $S_\alpha$. 
%We expect the
%conclusions not to depend qualitatively on these approximations in
%view of the fact that the effects we find involve distances $\kappa r
%\sim 5$--$10$.  {\tt I feel somewhat uncomfortable with this
%  statement: we say that the minimum depends sensitively on the
%  parameters and that the values of $\epsilon_{F,c}$ are close to the
%  limit of validity of the model. I have included a cautionary remark
%  in the conclusions, so we can perhaps drop this sentence
%  altogether.}
%The stress will be dominated by the charge density on the
%colloid--water surface, usually much larger than the one on the
%colloid--air surface:
%%Since the charge density
%%on the colloid--water surface is usually much larger than the one on the colloid--air surface
%%the former will dominate the stress which is exerted on the interface:%In the following, we treat
%the electrostatic problem by assuming that the total charge $q$ on the colloid--water
%surface is concentrated in the center of $S_\alpha$
%and by employing the Debye-H\"uckel approximation. Using the latter we find
%$p_{\rm osm} = (\epsilon_2/\,8\pi) \kappa^2\Phi_0^2$.
Then the osmotic pressure is given by $p_{\rm osm} =
(\epsilon_2/\,8\pi) \kappa^2\Phi_0^2$ and the electrostatic repulsion
between two colloids by $V_{\rm rep}(d)=q\Phi_0(d)$, where $\Phi_0$ is
the electrostatic potential at the interface. 
$V_{\rm rep}$ exhibits  a crossover from a screened Coulomb repulsion to a 
dipole repulsion at $\kappa d_{\rm c} \approx 8\dots 10$ \cite{Hur85}.
The single--colloid stress, Eq.~(\ref{eq:stress}), is 
dominated by $\Ep$ and $p_{\rm osm}$ for $\kappa r < 6 \dots 8$, yielding 
$\Pi(r) \propto \exp(-2\kappa r)/r^{4}$.
For larger distances, it is dominated by $\Ezp$ and the familiar dipole form arises, 
%$\Pi(r) \propto (\epsilon_1/\epsilon_2)r^{-6}$.
$\Pi(r) \propto r^{-6}$. The stress $\hat\Pi$ 
exerted by two colloids can again be determined by superimposing the
electric fields and potentials of the single--colloid solution, leading to
\bea
 \fl
 \label{eq:pim_hurd}
 \Pi_{\rm m} &=&
  \frac{\epsilon_1}{8\pi}\left( 1- \frac{\epsilon_1}{\epsilon_2}\right) (\Ezp)_1
  (\Ezp)_2   + 
% \\ \nonumber & &
   \frac{\epsilon_2-\epsilon_1}{8\pi} (\Ep)_1(\Ep)_2 \cos\phi + \frac{\epsilon_2}{8\pi}
  \kappa^2 (\Phi_0)_1(\Phi_0)_2\;.
\eea
As before, the subscript $\alpha=1,2$ denotes evaluation of the single--colloid  function at
$|\vect r-\vect r_\alpha|$ and $\cos\phi = 
(\vect r-\vect r_1)\cdot(\vect r-\vect r_2)/
|\vect r-\vect r_1||\vect r-\vect r_2|$. 
\begin{figure}
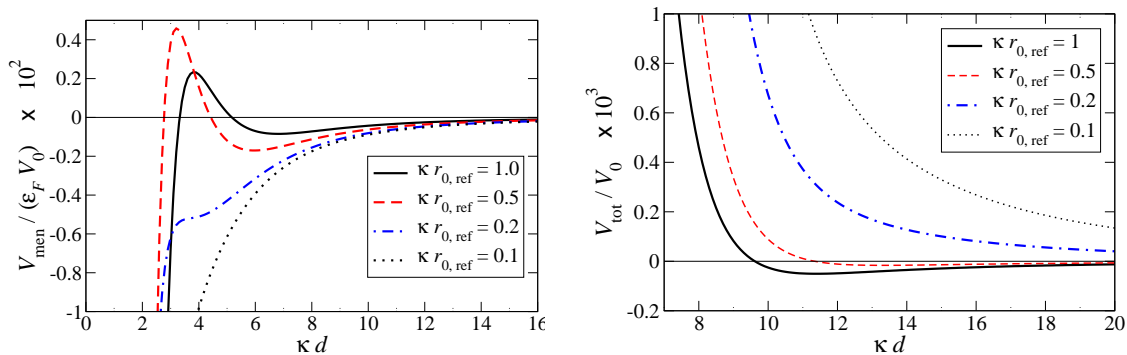

    \caption{\label{fig:vmen} (colour online) The capillary potential (left panel) obtained from the full numerical solution 
    in units of $V_0 = q^2 \kappa^2 \ro/\,2 \pi \epsilon_2$ for 
    $\kappa\ro=0.1 \dots 1.0$ and the
  parameters $\epsilon_2/\epsilon_1=81$ and $\kappa \lambda=100$. 
  The peak approximation fails for $\kappa d \alt 8$.
   The  total intercolloidal potential $V_{\rm tot}=V_{\rm men}+V_{\rm rep}$ (right panel) 
   for $\ef=0.6$ and the same parameters and units.
    }
 \vspace*{12mm}
  \begin{center}
    \epsfig{file=fig1.eps,width=7.2cm} \hspace{2mm}
    \epsfig{file=fig2.eps,width=7.2cm}
  \end{center}
\end{figure}
%Due to the aforementioned crossover 
%behavior it is 
%possible to perform a systematic $1/d$ expansion  for $u_{\rm m}$ as obtained from
%Eqs.~(\ref{eq:um}) and (\ref{eq:umb}) only for $d \gg d_{\rm c}$. 
%%Furthermore, such an expansion is complicated by the appearance
%%of the angle $\phi$ in Eq.~(\ref{eq:pim_hurd}). For the limit $d \gg d_{\rm c}$, 
%In this limit, the peak approximation
%of $\Pi_{\rm m}$ (not valid for intermediate $d$) is given by 
%\bea
% \Pi_{\rm m} &\approx& \sum_{\alpha=1,2} \left[
%  \frac{\epsilon_1}{8\pi}\left( 1- \frac{\epsilon_1}{\epsilon_2}\right) \Ezp(d) 
%  (\Ezp)_\alpha   + \right. \\ \nonumber
% & &  \frac{\epsilon_2-\epsilon_1}{16\pi} (\Ep'(d)+\Ep(d)/d) r_\alpha(\Ep)_\alpha  + \\ 
%  & &  \nonumber \left.\frac{\epsilon_2}{8\pi}
%  \kappa^2 \Phi_0(d)(\Phi_0)_\alpha\right]\;,\qquad r_\alpha = |\vect r-\vect r_\alpha| \;.
%\eea
%On this basis, we have calculated $u_{\rm m}$ and $V_{\rm men}$ 
% and we find for $d \gg d_{\rm c}$ that the capillary potential
%again decreases dipole--like $\propto d^{-3}$ as $V_{\rm rep}$ does.
In the limit $d\gg d_c$, $u_{\rm m}$ and $V_{\rm men}$ can be
estimated via an expansion in $1/d$ using a peak approximation for
$\Pi_{\rm m}$ as before. This yields an asymptotically attractive capillary potential 
which again decays $\propto d^{-3}$.
For intermediate separations $d$,
we have compared these  results with a full numerical solution to Eqs.~(\ref{eq:um})
and (\ref{eq:umb}) without any approximation for $\Pi_{\rm m}$ or $V_{\rm men}$. 
In Fig.~\ref{fig:vmen} (left panel) we show
the results for $V_{\rm men}$: For $d>d_{\rm c}$ the $1/d$--expansion and the 
numerical results agree with each other, both exhibiting the dipole--like 
asymptotic behavior. For 
smaller $d$ the extrapolated $1/d$--expansion predicts a repulsive capillary potential which is
borne out by the numerical results only partly as $\kappa\ro \to 1$. 
The superposition contribution $V_{\rm sup}$ is again negligible in the
expression for $V_{\rm men}$.
The total intercolloidal potential $V_{\rm tot}=V_{\rm men}+V_{\rm rep}$
is depicted in Fig.~\ref{fig:vmen} (right panel) for the exemplary value $\ef=0.6$.
%for which this model is still valid. 
 Asymptotically $V_{\rm tot}$ is attractive for $\ef > \varepsilon_{F,{\rm c}} 
\approx 0.34 [1+0.30/(\kappa\ro)^2]^{1/2}$,
obtained from a fit to numerical results for $\kappa \ro \in [0.1,2.0]$.

A quantitative comparison with experiments is difficult because the three important 
quantities $q$, $\theta$, and $\kappa^{-1}$, which enter into  $\ef$ and the scale of 
$V_{\rm men}$ have not been determined separately for the same system. 
Therefore, we can only estimate whether a minimum as obtained in Fig.~\ref{fig:vmen}
can occur in actual experiments.
For almost all experiments pure water is
claimed to have been used, for which $\kappa^{-1} \approx 1$ $\mu$m. 
The total charge 
is given by $q=2\pi\sigma\,R^2\,(1+\cos\theta)$ where $\sigma$
is the charge density.
%By direct observation,
%the contact angle was measured in Ref.~\cite{Ave00a} for polystyrene spheres
%to be $\theta\approx 70^o$ for water--oil interfaces and $\theta \le 30^o$ for
%water--air interfaces. 
Thus, for colloids with $R=0.5$ $\mu$m and $\theta=\pi/2$ at pure
water--air interfaces ($\gamma=0.07$ N/m), we obtain numerically $\ef \approx 45
(\sigma\,{\rm nm}^2/e)^2$. %with our approximation for the  stress~(\ref{eq:stress}). 
The potential scale at $T=300$ K is given by
$V_0=q^2\kappa^2\ro/(2\pi\epsilon_2) \approx 1.4 \cdot 10^8\,k_B
T(\sigma\,{\rm nm}^2/e)^2$. The value $\ef=0.6$ used in
Fig.~\ref{fig:vmen} corresponds to a charge density 
$\sigma=0.12$ $e/{\rm nm}^2$
(literature estimates for the actual charge density vary between 
$0.07$ \cite{Ghe97} and $0.53$
\cite{Sta00} $e/{\rm nm}^2$), in which case $V_0=2\cdot 10^6$ $k_B T$.
For $\kappa\ro=0.5$ the minimum in the 
total potential (see Fig.~\ref{fig:vmen}) occurs at
$\kappa d_{\rm min} \approx 13$ with $V_{\rm min} \approx -1.6\cdot 10^{-5} V_0=
32$ $k_B T$. This minimum is shallow enough that thermal movements of the colloids
around the minimum position should be visible, similar to  reports in the literature.

Such large charge densities actually call for solving the 
Poisson--Boltzmann equation near the colloids. Incorporating the correct geometry,
this is a very involved numerical task and will be considered in future work.
From that we expect the following modifications: The screening ions will concentrate near
the colloid in a layer with thickness of a few nm (as given by the Gouy--Chapman
length $l_{\rm G}=8\pi\epsilon_2/(\beta e \sigma)$). Outside this layer, 
the colloids appear as heavily screened objects with an effective charge
$q_{\rm eff}$  and the Debye--H\"uckel approximation is valid. This charge
$q_{\rm eff}$ enters both the electrostatic repulsion and the capillary attraction such
that asymptotically $V_{\rm rep} \propto q_{\rm eff}^2 (\ro/d)^3$ and
$V_{\rm men} \propto \ef\, q_{\rm eff}^2 (\ro/d)^3$. The total force
on one colloid (as expressed by $\ef$) is not determined by $q_{\rm eff}$,
but rather by the interfacial stress in the screening layer (dominated
by $p_{\rm os}$ and $\Ep$). Since this layer is thin compared to the colloid
radius, near the three--phase contact line the problem
appears to be similar to that of a charged plate half--immersed in water. For an order of
magnitude estimate of $\ef$ we have used the analytical solution
for a fully immersed plate and obtain 
$\ef \sim \epsilon_2/(\pi\gamma\beta^2 e^2 l_{\rm G})=1.2(\sigma\,{\rm nm}^2/e) $
for the parameters discussed above. Thus, $\ef=0.6$ (as used in 
Fig.~\ref{fig:vmen}) is obtained for $\sigma=0.5\,e/{\rm nm}^2$, not
too far from the Debye--H\"uckel result. We emphasize that
the occurrence of a potential minimum for $\ef>\varepsilon_{F,{\rm c}}$ (i.e., for a 
sufficiently large charge $q$) is a consequence of the
{\em intermediate distance} crossover in $\Pi$ 
 from being pressure--dominated (by $\Ep$ and $p_{\rm osm}$) to tension--dominated (by $\Ezp$).
This mechanism is captured correctly by the presently employed approximations
(point charge and Debye-H{\"u}ckel treatment).   

To summarize, we have calculated the effective intercolloidal potential 
for charged colloids floating on
a water interface. We have derived a general expression for the capillary potential
as an explicit functional of the stress on the interface. 
We have quantitatively studied the capillary potential when the stress is due to the electrostatic field of charged colloids in the cases that the
radius $R$ of the colloid compared with the Debye screening length
$\kappa^{-1}$ of 
water is either very large or of about the same size. In both cases the
asymptotic behavior of the capillary potential and of the direct  electrostatic 
repulsion is 
equal, $\propto d^{-3}$, but different in sign. 
The superposition approximation, predicting a capillary potential $\propto
d^{-6}$, is insufficient because it takes into account only the energy
change as the subsytem ``one colloid+surrounding meniscus" is shifted
vertically in its own (single--colloid) force field. The correction to the
superposition approximation, embodied in $\Pi_{\rm m}$, considers the
additional force by the second colloid.
The capillary attraction only 
dominates for a sufficiently large charge and  
the total intercolloidal
potential exhibits an attractive minimum only if $\kappa^{-1}\sim R$.
This minimum can be 
understood as a consequence of
the pressure--to--tension crossover in the stress acting on the interface.
%component of the electric field 
%near one colloid to a dominating normal field far away from the colloids.
The depth of the potential minimum is significantly reduced compared to the
natural energy scale $\gamma R^2$ of capillary interactions  and is of the
order of several $k_B T$ for parameters relevant for actual experimental
conditions. 
%
%One must be however cautious about the presence of a minimum, since
%this depends sensitively on the system parameters and occurs for
%values close to the theoretical limit of validity of the model.

\end{document}